\documentclass[reprint,superscriptaddress,amsmath,amssymb,aps,prl]{revtex4-2}
\pdfoutput=1

\usepackage{amsmath, amssymb,graphicx}
\usepackage{dcolumn}
\usepackage{bm}
\usepackage{graphicx}
\usepackage{bbm}
\usepackage{xcolor}
\usepackage{dutchcal}
\usepackage{physics}

\usepackage{hyperref}
\hypersetup{
  colorlinks,
  citecolor=magenta,
  linkcolor=magenta,
  urlcolor=magenta}

\def\be{\begin{equation}}
\def\ee{\end{equation}}

\def\nn{\nonumber}

\def\ra{\rangle}

\begin{document}

\title{Superresolution imaging with entanglement-enhanced telescopy}

\author{Isack Padilla}
\email{iacpad0795@arizona.edu}
\affiliation{College of Optical Sciences, University of Arizona, Tucson AZ 85721}

\author{Aqil Sajjad}
\email{asajjad@umd.edu}

\affiliation{College of Optical Sciences, University of Arizona, Tucson AZ 85721}
\affiliation{Department of Electrical and Computer Engineering, University of Maryland, College Park MD 20742}

\author{Babak N. Saif}
\email{babak.n.saif@nasa.gov}
\affiliation{NASA Goddard Space Flight Center, 8800 Greenbelt Rd, Greenbelt, MD 20771, USA}

\author{Saikat Guha}
\email{saikat@umd.edu}
\affiliation{College of Optical Sciences, University of Arizona, Tucson AZ 85721}
\affiliation{Department of Electrical and Computer Engineering, University of Maryland, College Park MD 20742}

\begin{abstract}
Long-baseline interferometry will be possible using pre-shared entanglement between two telescope sites to mimic the standard phase-scanning interferometer, but without physical beam combination. We show that spatial-mode sorting at each telescope, along with pre-shared entanglement, can be used to realize the most general multimode interferometry on light collected by any number of telescopes, enabling achieving quantitative-imaging performance at the ultimate limit pursuant to the baseline as afforded by quantum theory. We work out an explicit example involving two telescopes imaging two point sources.
\end{abstract}
\maketitle

\textit{Introduction}---Quantum entanglement shared over the future quantum internet~\cite{Wehner2018} could one day allow performing long-baseline interferometry over much larger distances than the few hundred meters currently possible for infrared or visible light with traditional techniques, without the need for physically transporting the light to a central location for interferometric beam combination~\cite{Gottesman2012,Tang2016,Khabiboulline2018,Khabiboulline2019, Chen2023, Czupryniak2023}. Additionally, tools from quantum information theory allow for a systematic quest for the {\em quantum optimal} optical-domain pre-processing and detection strategies for a given quantitative passive imaging problem, for any given telescope configuration~\cite{Sajjad2023}. For instance, the Cram\'er-Rao bound of classical estimation theory establishes the inverse of $N$ times the Classical Fisher Information (CFI) as a tight lower bound on the precision (variance) of estimating an unknown parameter from $N$ random-noisy observations. Given a measurement on $N$ copies of an information-bearing quantum state, the quantum Cram\'er-Rao bound establishes the Quantum Fisher Information (QFI) as an upper bound on the CFI attainable by any physically allowed measurement on the quantum states~\cite{Helstrom1976}. This allows searching for quantum-optimal measurement strategies. For example, in the context of a single-telescope, sorting the collected light in an optimal spatial mode basis prior to photon detection can allow for a far-higher precision information extraction compared to the standard receiver strategy of direct detection of the light's intensity profile on a pixelated imaging screen. A stark example of this classical-quantum separation appears in the task of resolving two equally bright point sources and estimating their separation, where the CFI for image-plane direct detection falls to zero as the separation approaches zero, reflecting Rayleigh's curse~\cite{LordRayleigh1879}. However, the QFI is a non-zero constant that does not approach zero even in the small separation limit, and can be attained by measuring the incoming photons in the Hermite Gauss (HG) basis for a Gaussian aperture~\cite{Tsang2016b}. These gains persist in the face of more complex problems such as imaging extended objects~\cite{Dutton2019}, classifying objects from a library~\cite{Grace2022}, and localizing multiple point sources~\cite{Lee2022}, all in the traditionally-unresolvable sub-Rayleigh regime.

The same paradigm is also applicable to long-baseline interferometry, where we can combine the light collected by several telescopes in an optimal basis for measurement~\cite{Cosmo2020, Wang2021, Bojer2022, Sajjad2023}. The most general $n$-telescope system based on this framework would involve a mode sorter collecting $K$ spatial modes at each site, which would be brought to a central location through single-mode fibers and combined in a linear interferometer whose outputs we measure for photon clicks~\cite{Sajjad2024}. The unitary mixing matrix of this interferometer between the $nK$ input modes and the same number of output modes, would then be optimally chosen for our specific quantitative imaging task based on quantum estimation theory techniques. Current long-baseline interferometry systems do not yet incorporate this perspective and are limited to only interfering signals from two telescopes at a time, and scanning for the phase difference. For example, if we have three telescopes $T_1$, $T_2$ and $T_3$, we interfere the light collected by $T_1$ with $T_2$, $T_1$ with $T_3$, and $T_2$ with $T_3$, but not all of them together. This is because combining light from more than a pair of telescopes exacerbates the already-difficult phase stabilization needed for optical-frequency long-baseline imaging. Furthermore, existing literature on quantum entanglement-assisted long-baseline interferometry~\cite{Gottesman2012,Tang2016,Khabiboulline2018,Khabiboulline2019, Chen2023, Czupryniak2023} focuses only on mimicking the standard phase scanning technique for combining two telescopes. 

\begin{figure}
    \centering
    \includegraphics[width=1\linewidth]{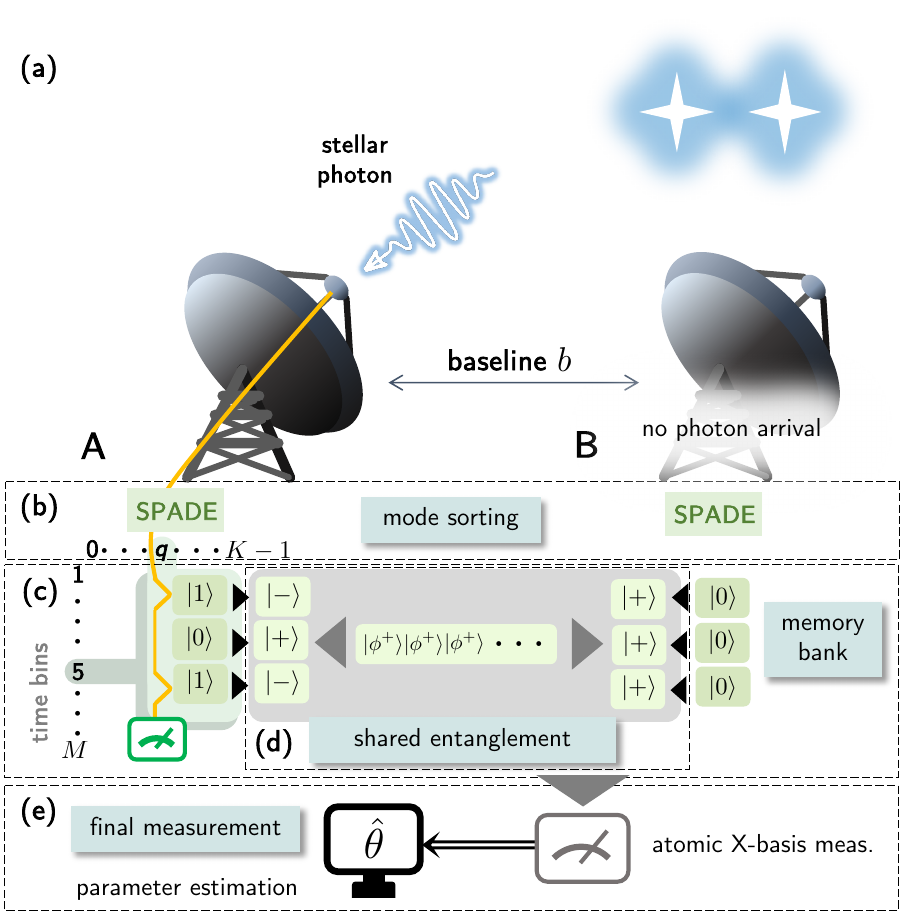}
    \caption{(a) A two-telescope array of baseline $b$ points toward two weakly emitting stars of angular separation $2\theta$. A star photon arriving at site A is shown. (b) The incoming photon is fed into a spatial mode demultiplexer (SPADE). An excitation is shown to occur in the second mode basis and in the fifth time bin of a block of integration time in which roughly one photon arrives. (c) The photonic state is loaded into the memory qubits via photon-memory CNOT gates, a compressive binary encoding, and performing X-basis measurements on the photon. (d) Entangled pairs pre-distributed among the telescope sites assist in performing a sequence of operations that reveal the arrival time and spatial mode index, which combined with (e) The X measurement results of relevant memory atoms, resuls in a single-bit post-processed outcome whose empirical probability over measurements of many time blocks containing one photon each, is the sufficient statistic to estimate $\theta$ at the QFI-mandated precision limit.}
    \label{fig:protocol}
\end{figure}

In this paper (also see~\cite{Padilla2024}), we begin with an explicit construction of the QFI-achieving strategy using shared entanglement between two telescopes for estimating the separation between two point sources. We next describe how to generalize this strategy to use pre-distributed entanglement among $n$-telescope sites, with each telescope equipped with a spatial mode sorter, to realize simultaneous beam combination from multiple telescopes in a general interferometer. This thus allows harnessing the full power of optimization techniques from quantum estimation theory. Our overall protocol is summarized in Fig.~\ref{fig:protocol}.

\textit{Two-telescope set-up}---Consider two telescopes $A$ and $B$ at aperture-plane positions $y_A=-\beta$ and $y_B=\beta$ respectively, labeled with index $\alpha\in\{A,B\}$. The quantum state of a single photon originating from a point source at $x_s$ in the object plane, where $s=1,\ldots, n_e$ is an index labeling the $n_e$ points the scene is composed of, is a superposition of starlight collected by the two telescopes: 
\begin{align}
\left|\psi^{(s)} _{AB}\right\rangle = \frac{1}{\sqrt{2}} \left(\left|\psi_{A} ^{(s)}\right\rangle + \left|\psi_{B} ^{(s)}\right\rangle\right),
    \label{psiab}
\end{align}
where $\left|\psi_{\alpha}^{(s)}\right\rangle$ is the state of a single photon originating at source $s$ collected by telescope $\alpha\in\{A,B\}$. Consider spatial mode functions $\phi_q(x)$ with $q=0,1,\ldots,\infty$, with $x$ the imaging screen position coordinate for a single aperture at the aperture plane origin $y=0$. Let us define $|0_{\alpha_{q}}\rangle$ as the vacuum state of the $q$-th mode.
Let $|0_\alpha\rangle \equiv \bigotimes_{i=0}^{\infty}  |0_{\alpha_{i }}\rangle$
be the vacuum in all spatial modes at site $\alpha$.
We then have the states $|1_{A_q},0_{B}\rangle\equiv a_{A_q}^{\dagger}|0_A,0_B\rangle$,
and
$|0_{A},1_{B_q}\rangle \equiv a_{B_q}^{\dagger}|0_A,0_B\rangle$
for a single photon in mode $q$ of the telescope at site $A$, and 
$B$, respectively,
with $a_{\alpha_q}^{\dagger}$ the creation operator for a photon in the $q$-th mode at site $\alpha$.
We will refer to these states as the {\em local spatial modes} of the two telescopes.

Expressing \eqref{psiab} in terms of these local modes and assuming that we are extracting the first $K$ modes at each site and that the final measurement step will herald a photon arrival, we can work with the projected state:
{\small\begin{align}
\left|\psi^{(s)} _{AB}\right\rangle \to \left|\psi^{(s)} _{AB, K}\right\rangle &\equiv  \frac{1}{\sqrt{2}}\sum_{q=0}^{K-1}\eta_q(x_s)\left( e^{-i \beta x_s} |1_{A_q},0_{B}\rangle \right. \nn \\  & \quad\quad\quad\quad\quad\quad\left.+ e^{i \beta x_s} |0_{A},1_{B_q}\rangle\right).
\label{psiab3}
\end{align}}
Here the exponentials arise from the apertures being located at $y_A=-\beta$ and $y_B=\beta$ instead of the aperture plane origin~\cite{Sajjad2023}. The coefficients
{\small $\eta_q(x_s) \equiv \Gamma_q(x_s)/\sqrt{\sum_{r=0}^{K-1} \Gamma_r ^2(x_s)}
$} where {\small$\Gamma_q(x_s) = \int_{-\infty}^\infty \text{d}x \phi_q^* (x) \psi(x-x_s)$} is the correlation function between the single aperture point spread function (PSF) $\psi(x)$ and the mode functions $\phi_q(x)$ for a single telescope located at the aperture plane origin.
The denominator arises from the projection onto the $0\leq q\leq K-1$ subspace, and approaches unity for large $K$, so we recover $\eta_q(x_s) \to \Gamma_q(x_s)$ 
 when $K$ approaches infinity.
 
Our goal is to measure the joint quantum state of the light collected by telescopes $A$ and $B$ for a general scene composed of $n_e$ point sources at positions $s = 1, \ldots, n_e$ (which we describe in the next section) in the basis:
\begin{align}
\left|\phi_{AB,q} ^\pm\right\rangle =\frac{1}{\sqrt{2}}\left(
|1_{A_q},0_{B}\rangle \pm |0_{A},1_{B_q}\rangle\right),
\quad q=0,\ldots,\infty.
   \label{phipm_original}
\end{align}
This measurement---different from the conventional phase-scanning approach---corresponds to interfering mode indices $q$ from the two telescopes pairwise in 50-50 beamsplitters followed by photon detection on both outputs, for $q = 0, \ldots, K-1$. This was shown to achieve the QFI for estimating the separation between two equally-bright point sources by a two-telescope system~\cite{Sajjad2023}.

\textit{The density operator}---Following the weak source model of~\cite{Tsang2016b}, we assume a small average number of photons arriving in a given temporal mode $\epsilon\ll1$. The incoming photon density matrix for a single temporal mode
\begin{align}
    \rho=(1-\epsilon)\rho_{0,AB} + \epsilon\sum_{s=1}^{n_e} b_s \left|\psi^{(s)} _{AB, K}\right\rangle\left\langle\psi^{(s)} _{AB, K}\right| \,+O(\epsilon^2),
\end{align}
where $\rho_{0,AB}=|0_A,0_B\rangle\langle 0_A, 0_B|$, the coefficients $b_s$ denote the relative brightnesses of the individual point sources in the scene satisfying $\sum_{s=1}^{n_e} b_s=1$. Across $M$ temporal modes, the density matrix of the collected light is:
{\small\begin{align}
\rho^{\otimes M}_{AB} = (1-M\epsilon)\rho_{0,AB}^{\otimes M}
+\epsilon \sum_{s=1}^{n_e} b_s \left|\psi^{(s)} _{AB,K,m}\right\rangle\left\langle \psi^{(s)} _{AB,K,m} \right| \,+O(\epsilon^2). 
\label{rhoabM}
\end{align}}
\noindent Here, $m\in\{1,\ldots,M\}$ indexes the temporal modes, and
{\begin{align}
\left|\psi^{(s)} _{AB,K,m} \right\rangle =\dfrac{1}{\sqrt{2}}\sum_{q=0}^{K-1} \eta_q (x_s) &\left(e^{i\beta x_s}\left|\widetilde0_{A},\widetilde1_{B_{mq}}\right\rangle\nn
\right.\\&\left.+ e^{-i\beta x_s}\left|\widetilde1_{A_{mq}},\widetilde0_{B}\right\rangle \right),
\end{align}}
where {\small$|\widetilde 0_\alpha\rangle \equiv \bigotimes_{i=0}^{K-1} \bigotimes_{j=1}^M |0_{\alpha_{ji}}\rangle$} is the vacuum for all the $KM$ spatio-temporal modes at site $\alpha$, and {\small$|\widetilde 1_{\alpha_{mq}}\rangle = a_{\alpha_{mq}} ^\dagger |\widetilde 0_\alpha\rangle$} is a single photon in the spatio-temporal mode $(m,q)$ at site $\alpha$, and vacuum elsewhere.

\textit{Encoding into memory qubits}---Instead of mapping each spatio-temporal photonic mode  onto a quantum memory, we apply the logarithmic compression proposed in~\cite{Khabiboulline2018,Khabiboulline2019}.
For each spatial mode at each site, we introduce $\overline{M}=\log_2(M+1)$ memory qubits initialized in $|0\rangle$ to capture the temporal mode of an incoming photon.
Thus, we have $\overline{M}K$ memory qubits at each site described by memory-registers $\overline{A}$ and $\overline{B}$ at sites $A$ and $B$, respectively, labeled $\overline{\alpha}\in\{\overline{A},\overline{B}\}$.

We introduce indices $k\in\{1,\dots,\overline{M}\}$, $i\in\{0,\dots,K-1\}$ and $j\in\{1,\dots,M\}$ for the photonic and memory qubits. Here, $i=q$ and $j=m$ correspond to the spatio-temporal mode pair $(m, q)$ where the photonic excitation is present. The initial state of the memory qubits at each site is thus {$\left|0_{\overline\alpha}\right\rangle\equiv\bigotimes_{i=0}^{K-1}\bigotimes_{k=1}^{\overline{M}}|0_{\overline{\alpha}_{ki}}\rangle$}.

Taking each spatio-temporal photonic mode as encoding a single-rail qubit where the presence or absence of a photon encodes qubit states $|0\ra$ or $|1\ra$, respectively, we apply a collection of CNOT gates from these photonic qubits to the memory ones at each site such that for an incoming photon in temporal mode $m$, a subset of the memory qubits are flipped to $|1\rangle$, corresponding to the binary encoding of $m$. We accomplish this by applying at each site the unitary operation,
{\small
\begin{equation}
{U}_{\alpha\overline\alpha}=\bigotimes_{i=0}^{K-1}\bigotimes_{j=1}^{M}\bigotimes_{k=1}^{\overline{M}}{({U_{\alpha_{ji}\overline{\alpha}_{ki}}})^{w_{kj}}},\quad\alpha\in\{A,B\},
\label{CNOT-at-each-site}
\end{equation}}
where ${U_{\alpha_{ji}\overline{\alpha}_{ki}}}$ denotes a CNOT gate from the photonic qubit corresponding to spatial mode $i$ and temporal mode $j$ at site $\alpha$
to the memory qubit number $k$ associated with spatial mode $i$ at the same location. The symbol $w_{kj}$ is the $k$-th digit in the binary representation of $j$.

The action of the unitary in Eq.~\eqref{CNOT-at-each-site} thus gives {\small${U}_{\alpha\overline\alpha}
|\widetilde1_{\alpha_{mq}}\rangle|0_{\overline \alpha}\rangle
= |\widetilde1_{\alpha_{mq}}\rangle|1_{\overline \alpha_{mq}}\rangle
$} where  {\small$|1_{\overline\alpha_{ m q}}\rangle
\equiv \bigotimes_{i=0}^{K-1}\bigotimes_{k=1}^{\overline{M}} X_{\overline\alpha_{i k}} ^{\delta_{i q} w_{km}} |0\rangle_{\overline{\alpha}_{ki}}$}
is the logical representation of the mapped excitation in the memory. Here $\delta_{i q}$ is the kronecker delta, and $X_{\overline\alpha_{ik}}$ denotes the Pauli $X$ operator acting on the memory qubit of register $\overline\alpha$ corresponding to the $k$th memory qubit of spatial mode $i$, flipping it from $|0\ra$ into $|1\ra$. Upon applying the CNOTs at both sites, the joint state of the photonic modes and the quantum memories becomes: 
{\small\begin{align}
    \left| \psi_{AB\overline{AB},m}^{(s)} \right\rangle \equiv &\dfrac{1}{\sqrt{2}}\sum_{q=0}^{K-1} \eta_q (x_s) \left(e^{i\beta x_s}\left|\widetilde0_{A},\widetilde1_{B_{mq}}\right\rangle\left|0_{\overline A},1_{\overline B_{mq}}\right\rangle \right. \nonumber\\
    & \left. + e^{-i\beta x_s}\left|\widetilde1_{A_{mq}},\widetilde0_{B}\right\rangle \left|1_{\overline A_{mq}},0_{\overline B}\right\rangle \right).
    \label{psiabab}
\end{align}
}

Next, we disentangle the photon-memory system by measuring the photonic qubits in the single-rail $X$ basis {\small $|\pm\rangle=(|0\rangle\pm|1\rangle)/\sqrt 2$}.
Expressing  the photonic qubits for spatio-temporal mode $(m, q)$ in the $|\pm\ra$ basis, we rewrite the photon-memory joint state \eqref{psiabab} as:
\begin{widetext}
\begin{align}
    \left| \psi_{AB\overline{AB},m}^{(s)} \right\rangle = &   \frac{1}{2}\sum_{q=0}^{K-1}\eta_q(x_s)
\left(\left|+_{A_{mq}},+_{B_{mq}}\right\rangle-\left|-_{A_{mq}},-_{B_{mq}}\right\rangle\right)
\left(e^{i\beta x_s}\left|0_{\overline A},1_{\overline B_{mq}}\right\rangle
+e^{-i\beta x_s}\left|1_{\overline A_{mq}},0_{\overline B}\right\rangle\right)  \nonumber \\ 
&  \quad + \frac{1}{\sqrt{2}}\sum_{q=0}^{K-1}\eta_q(x_s)
\left(\left|-_{A_{mq}},+_{B_{mq}}\right\rangle-\left|+_{A_{mq}},-_{B_{mq}}\right\rangle\right)
\left(e^{i\beta x_s}\left|0_{\overline A},1_{\overline B_{mq}}\right\rangle-e^{-i\beta x_s}\left|1_{\overline A_{mq}},0_{\overline B}\right\rangle\right).
\label{photon-memory-state-even-odd-expansion}
\end{align}
\end{widetext}
Here $|\pm_{\alpha_{mq}}\rangle$ is the photonic state with the spatio-temporal mode pair $(m, q)$ at site $\alpha$ in $|\pm\rangle$ and the remaining photonic qubits at both sites in $|0\rangle$.
The  memory state after measuring the photons is
{\small\begin{align}
\left|\chi_{\overline{AB}, m}^{(s)}\right\rangle = \dfrac{1}{\sqrt{2}}\sum_{q=0}^{K-1}&\eta_q(x_s) \left(e^{i\beta x_s }  \left|0_{\overline A},1_{\overline B_{mq}}\right\rangle  
    \right. \nonumber \\ & \left.
    +f_{mq}e^{-i\beta x_s }\left|1_{\overline A_{mq}},0_{\overline B}\right\rangle\right),
    \label{memory-ket}
\end{align}}
where $f_{mq}$ is $1$ (or, $-1$) if we obtain the same (or, opposite) results $|\pm_{A_{mq}}, \pm_{B_{mq}}\ra$ (or, $|\pm_{A_{mq}}, \mp_{B_{mq}}\ra$) in the spatio-temporal mode pair $(m, q)$ at both sites. However we will not know whether we have $1$ or $-1$ until we determine the spatio-temporal mode of the incoming photon in the decoding stage (discussed next) and therefore will store all the $X$ basis measurement results for later use.

\textit{Decoding}---To determine the spatio-temporal mode of the incoming photon, for each $k \in \left\{0, \overline{M}\right\}$ and $i \in \left\{0, K-1\right\}$ labeling a pair of memory qubits at the two sites, we introduce the Bell state
{\small$| \phi^+_{C_{ki}D_{ki}} \rangle=(|0_{C_{ki}} 0_{D_{ki}}\rangle+|1_{C_{ki}} 1_{D_{ki}}\rangle)/\sqrt{2}$}
with memory qubits $C_{ki}$ and $D_{ki}$ at sites $A$ and $B$, respectively.

We then apply a pair of CZ operations
{\small $V_{\overline{A}_{ki}\overline{B}_{ki} C_{ki}D_{ki}} \equiv
CZ_{\overline{A}_{ki} C_{ki}} \,CZ_{\overline{B}_{ki} D_{ki}}$},
between each memory qubit at each site and the accompanying local qubit of the ancillary bell pair.
Note that
\begin{equation}
V_{\overline{A}_{ki}\overline{B}_{ki} C_{ki}D_{ki}} \left|0_{\overline{A}_{ki}} 1_{\overline{B}_{ki}}\right\rangle \left|\phi^+ _{C_{ki} D_{ki}}\right\rangle = \left|0_{\overline{A}_{ki}} 1_{\overline{B}_{ki}}\right\rangle
\left|\phi^- _{C_{ki} D_{ki}}\right\rangle \nonumber,
\label{vv1}
\end{equation}
\begin{equation}
V_{\overline{A}_{ki}\overline{B}_{ki} C_{ki}D_{ki}} \left|1_{\overline{A}_{ki}} 0_{\overline{B}_{ki}}\right\rangle \left|\phi^+ _{C_{ki} D_{ki}}\right\rangle
= \left|1_{\overline{A}_{ki}} 0_{\overline{B}_{ki}}\right\rangle
\left|\phi^- _{C_{ki}   D_{ki}}\right\rangle \nonumber,
    \label{vv2}
\end{equation}
\begin{equation}
V_{\overline{A}_{ki}\overline{B}_{ki} C_{ki}D_{ki}} \left|0_{\overline{A}_{ki}} 0_{\overline{B}_{ki}}\right\rangle
\left|\phi^+ _{C_{ki} D_{ki}}\right\rangle 
= \left|0_{\overline{A}_{ki}} 0_{\overline{B}_{ki}}\right\rangle \left|\phi^+ _{C_{ki} D_{ki}}\right\rangle \nonumber,
    \label{vv3}
\end{equation}
where {\small$| \phi^-_{C_{ki}D_{ki}}\rangle=(|0_{C_{ki}} 0_{D_{ki}}\rangle-|1_{C_{ki}} 1_{D_{ki}}\rangle)/\sqrt{2} \,=\left(|+_{C_{ki}} -_{D_{ki}}\rangle +|-_{C_{ki}} +_{D_{ki}}\rangle\right)/\sqrt{2}$}.
Thus, the Bell states flip when there is an excitation at one of the two sites.

We can determine whether a Bell state flipped or not by measuring each of its qubits in the $X$ basis since 
{\small $|\phi^+_{C_{ki}D_{ki}}\rangle=\left(|+_{C_{ki}} +_{D_{ki}}\rangle +|-_{C_{ki}} -_{D_{ki}}\rangle\right)/\sqrt{2}$}. Thus, even parity results $\left|\pm_{C_{ki}}, \pm_{D_{ki}}\right\rangle$ mean an unflipped Bell pair, and odd parity results $|\pm_{C_{ki}}, \mp_{D_{ki}}\rangle$ signify a flipped one. This tells us which memory qubits have the excitations, revealing the temporal mode $m$ in which the photon arrived, and also serving as a projective measurement to determine its spatial mode $q$ with outcome probabilities (conditional on there being an incoming photon) $p_{q}(\theta)=\eta^2_{q}(\theta)$. We now revert to our stored results of the $X$-basis measurements on the photonic qubits to find $f_{mq}$, which is $1$ and $-1$ for even and odd parity results, respectively. The remaining task therefore is to determine if the photon was in the symmetric or the anti-symmetric combination of mode $q$ collected at the two sites (to mimic the pairwise receiver of~\cite{Sajjad2023}). 

Now, the memory qubits accompanying the unflipped Bell states are all in $\left|0\right\rangle$ and can be dropped going forward.
Let $N_m\in\{1,...,\overline{M}\}$ be the number of Bell pairs found to have flipped into $\left| \phi^- \right\rangle$.
We now define subsystems $E$ and $F$ containing only the $N_m$ remaining memory qubits of registers ${\bar A}$ and ${\bar B}$, respectively, labeled as $\gamma\in\{E,F\}$. Then we have the reduced state consisting of $2N_m$ entangled qubits
{\small$\sum_{s=1}^{n_e} b_s \left| \omega_{EF,mq}^{(s)}\right\rangle\left\langle \omega_{EF,mq}^{(s)}\right|$},
where
{\small \begin{align}
    \left| \omega_{EF,mq}^{(s)}\right\rangle = \frac{1}{\sqrt 2}&\left( e^{i\beta x_s}|\textbf{0}_{E_{mq}},\textbf{1}_{F_{mq}}\rangle\right. \nonumber \\
    & \left.
     + f_{mq} e^{-i\beta x_s}|\textbf{1}_{E_{mq}},\textbf{0}_{F_{mq}}\rangle\right),
\label{memket}
\end{align}}
Here {\small $\left|\textbf{0}_{\gamma_{mq}}\right\rangle \equiv \bigotimes_{\mu=1}^{N_m} |0_{\gamma_{\mu m q}}\rangle$} and {\small $\left|\textbf{1}_{\gamma_{mq}}\right\rangle = \bigotimes_{\mu=1}^{N_m} |1_{\gamma_{\mu m q}}\rangle$} with $\gamma\in\{E,F\}$.

We should mention that we can also have a more general version of the above heralding procedure with GHZ instead of Bell states, such as one that only determines the temporal mode $m$ without necessarily collapsing into a single spatial mode. We can then measure the surviving memories in an arbitrary basis to mimic an arbitrary linear interferometric measurement on the spatial modes collected across the two sites, which may be needed for QFI-optimal measurements for more general quantitative imaging tasks.

To determine if the incoming photon was in the symmetric or the anti-symmetric combination of the $q$-th mode signals collected at the two sites, i.e., a measurement of the erstwhile photonic state in the $|\phi_{AB,q} ^\pm\rangle$ basis, we need to measure the memory state~\eqref{memket} in the basis:
\begin{align}
     \left|\zeta^\pm_{EF,mq}\right\rangle =\frac{1}{\sqrt{2}} (|\textbf{0}_{E_{mq}},\textbf{1}_{F_{mq}}\rangle \pm |\textbf{1}_{E_{mq}},\textbf{0}_{F_{mq}}\rangle),
     \label{zetapm}
 \end{align}
where the measurement outcome needs to be flipped if $f_{mq} =-1$. The $\left|\zeta^\pm_{EF,mq}\right\rangle$ basis measurement is realized by measuring each of the $2N_m$ qubits contained in the $E$ and $F$ registers in the $X$ basis and considering the $N_m$ parities among the corresponding qubit pairs in the two registers. An even number of odd parity $X$-measurement results ($|\pm_{E_{\mu m q}},\mp_{F_{\mu m q}}\rangle$, $1 \le \mu \le N_e$) implies the $|\zeta^+ _{EF,mq}\rangle$ outcome, and an odd number of odd-parity results implies the $|\zeta^- _{EF,mq}\rangle$ outcome~\cite{Padilla2024}.

Writing \eqref{memket} in terms of $|\zeta^\pm_{EF,mq}\rangle$ and putting together the various pieces, we obtain the conditional probabilities $p_+(x_s) = \cos^2(\beta x_s)$ and $p_-(x_s) = \sin^2(\beta x_s)$ that given an incoming photon is in spatial mode $q$, it is in $|\phi_{AB,q}^\pm\rangle$. Overall, we obtain the full probabilities for a photon originating from a source at $x_s$ to be in the modes $|\phi_{AB,q}^\pm\rangle$: 
\begin{align}
    P_{q+}(x_s) = p_+(x_s) P_q(x_s)= \cos^2 (\beta x_s) \eta_q ^2(x_s),
\end{align}
\begin{align}
    P_{q-}(x_s) = p_-(x_s) P_q(x_s)=\sin^2 (\beta x_s) \eta_q ^2(x_s).
\end{align}

\begin{figure}
    \centering
    \includegraphics[width=1\linewidth]{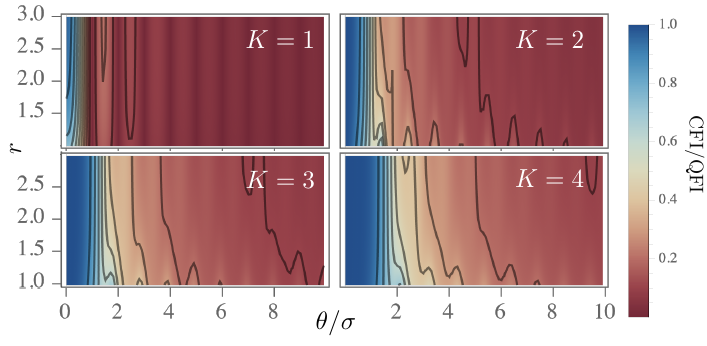}
    \caption{CFI normalized to the QFI, plotted as a color chart, versus separation $\theta/\sigma$ and baseline to aperture-diameter ratio $r$. Four values of the spatial-mode cuffoff $K$ are shown, with the top right corresponding to a binary SPADE ($K=2$) attaining the QFI in the sub-Rayleigh regime ($\theta/\sigma < 1$).}
    \label{fig:results}
\end{figure}

\textit{Attaining the quantum limit of two-point separation estimation}---Let us consider the problem of estimating $\theta$, the half-angular separation between two point sources located at $x_s \in \left\{-\theta, \theta\right\}$, $s = 1, 2$. We will consider a one-dimensional setup with two length-$\delta$ telescopes (i.e., sinc-function PSFs) centered at $y_A = -\beta$ and $y_B = \beta$. The QFI for estimating $\theta$ was found to be $4 \pi^2 N(3r^2+1)/3\sigma^2$~\cite{Sajjad2023}, where $N$ is the total number of photons collected across both telescopes, $\sigma=2\pi/\delta$ is the single-telescope Rayleigh separation angle, and $r=2\beta/\delta$ is the ratio of the mid-point separation between the two telescopes to the individual aperture size.

In the $K\to\infty$ limit where we collect a large number of spatial modes, $\eta_q(x_s) \to \Gamma_q(x_s)$, and we recover the probabilities obtained in~\cite{Sajjad2023} for measuring the joint state of the photonic state across the two telescopes in the pairwise basis~\eqref{phipm_original}, but using shared entanglement. It was shown in~\cite{Sajjad2023} that the CFI for the pairwise measurement for estimating the separation between two equally bright points, is equal to the QFI, assuming prior knowledge of the centroid with the telescopes perfectly pointed at it~\cite{Tsang2016b, Grace2020c, Sajjad2021}.
In Fig.~\ref{fig:results}, we show the ratio of the CFI to this QFI as a function of the angular separation $\theta/\sigma$ and $r$ for different $K$ values. Notice that, even when taking the fundamental mode only (i.e. $K=0)$, the QFI can still be saturated in the sub-Rayleigh region of interest.
        
\textit{Generalizing to arbitrary measurements on any number of telescopes}---Consider $n$ telescopes each collecting $K$ spatial modes.
We can bring all the signals to a central location through single-mode fibers, and feed them into a linear interferometer which mixes them in some basis of the $nK$ modes collected across the various sites~\cite{Sajjad2023}, thus measuring the incoming photons in the said basis. We can create an arbitrary linear interferometer with $nK$ inputs and the same number of outputs by employing $nK(nK-1)$ 50-50 beam splitters and the same number of phase shifters~\cite{Clements2016}. To carry this out with shared entanglement, we need the following ingredients:
\begin{enumerate}
\item A multi-site generalization of reading the parity of the $X$ basis measurement results of the photonic qubits and incorporating them into our protocol. This is a somewhat simple book-keeping exercise in principle, but becomes a bit more elaborate than just considering the parity of the $X$ basis results associated with a single spatio-temporal mode.
\item Once the state of the photons from the scene has been loaded on to the quantum memories, an $n$-site and $K$-spatial mode generalization of our heralding procedure for determining the spatio-temporal mode of an incoming photon with ancilla Bell states described in the Decoding section. This can be accomplished by employing a GHZ state in place of a Bell state to determine the temporal mode, and not necessarily collapsing the state to a single spatial mode to allow mixing multiple spatial modes. For more details, see~\cite{Padilla2024}.
\item The ability to perform on a pair of logical qubits comprising of several atomic memory qubits, an operation that mimics the action of a 50-50 beam splitter between two photonic modes restricted to the space spanned by a single photon across the two modes. The two logical qubits may be at the same site or at distant telescope locations. Specifically, the 50-50 beam splitter action on a generic one-photon state across two optical modes $A$ and $B$ is:
\begin{align}
a |1_A 0_B\rangle & + b |0_A 1_B\rangle 
\to \frac{a+b}{\sqrt{2}} |1_{A'} 0_{B'}\rangle 
+ \frac{a-b}{\sqrt{2}}|0_{A'} 1_{B'}\rangle,
\end{align}
where $A'$ and $B'$ are the output modes.
Then, if the states of the input optical modes are transferred on to logical qubits $\overline{E}$ and $\overline{F}$, the equivalent operation on these logical qubits can be accomplished with three steps: a CNOT from logical qubit $\overline{E}$ to $\overline{F}$, a Hadamard operation on qubit $\overline{E}$, and a CNOT from qubit $\overline{E}$ to $\overline{F}$.
When the two logical qubits are at distant locations, the CNOTs need to be implemented using gate teleportation~\cite{gottesman1999quantum} by using a shared Bell pair.
\item The ability to apply any arbitrary phase shift to an optical mode or a single-qubit phase on an atomic qubit once the photonic state has been loaded on to quantum memories. This is straightforward to implement in either domain.
\item (Optional) An $n$ site and $K$-spatial-mode generalization of the Clements {\em et al.} approach~\cite{Clements2016} for implementing the same unitary transformation more efficiently with fewer steps by employing pre-shared GHZ states instead of only using Bell states and mimicking 50-50 beam splitters.
\end{enumerate}
We thus have the ingredients needed to generalize our protocol to arbitrary measurements on any number of telescopes.

\textit{Conclusion}---Shared entanglement offers a potentially promising method for baseline interferometry over substantially larger distances than the current range of a hundred or so meters possible with existing technology. However, this topic still remains mostly unexplored, with a few pioneering works focusing on imitating the phase-scanning interferometer involving two distant telescopes~\cite{Gottesman2012, Khabiboulline2018}.
Here we have presented a framework for addressing a general imaging or parameter estimation task by measuring the light entering a multiple-telescope system in any arbitrary basis by employing entanglement. We have explicitly illustrated how this can be done by building an entanglement-based equivalent of a two-telescope system where we collect $K$ spatial modes at each site, and for each mode bring the light to a central location, combine it in a 50-50 beam splitter, and measure the outputs. Such a receiver measures the incoming photons in terms of the sum and difference of the signals for each spatial mode from the two locations.
We have then described how this approach can be generalized to carry out any arbitrary measurement on an $n$ telescope system, which can be selected based on optimization techniques from quantum estimation theory for the imaging or parameter estimation task at hand. We hope our work will inspire more research on the subject and help pave the way towards a practical implementation of baseline interferometry with shared entanglement.  

The authors thank Michael Grace, Gabe Richardson, Brittany McClinton, Jayadev Rajagopal, Ryan Lau, and Stephen Ridgway for helpful discussions, and Prajit Dhara for his comments on the manuscript. This work was partially funded by NASA grant number 80NSSC22K1030 and AFOSR grant number FA9550-22-1-0180.

\end{document}